\documentclass{elsart}

\usepackage {graphicx}
\newcommand{\D}{{\rm d}}

\newcommand{\PP}{\mbox{\boldmath$P$}}

\begin{document}

\begin{frontmatter}

\title{\textbf{Order--disorder separation: Geometric revision}}

\author{Alexander Gorban\corauthref{cor1}}
\ead{ag153@le.ac.uk}
\corauth[cor1]{Corresponding author: Centre
for Mathematical Modelling, University of Leicester, University
Road, Leicester, LE1 7RH,  UK}
\address{University of Leicester,
Leicester,  UK}


\maketitle

\begin{abstract}
After Boltzmann and Gibbs, the notion of disorder in statistical
physics relates to ensembles, not to individual states. This
disorder is measured by the logarithm of ensemble volume, the
entropy. But recent results about measure concentration effects in
analysis and geometry allow us to return from the ensemble--based
point of view to a state--based one, at least, partially. In this
paper, the order--disorder problem is represented as a problem of
relation between distance and measure. The effect of strong
order--disorder separation for multiparticle systems  is described:
the phase space could be divided into two subsets, one of them (set
of disordered states) has almost zero diameter, the second one has
almost zero measure. The symmetry with respect to permutations of
particles is responsible for this type of concentration. Dynamics of
systems with strong order--disorder separation has high average
acceleration squared, which can be interpreted as evolution through
a series of collisions (acceleration--dominated dynamics). The time
arrow direction from order to disorder follows from the strong
order--disorder separation. But, inverse, for systems in space of
symmetric configurations with ``sticky boundaries" the way back from
disorder to order is typical (Natural selection). Recommendations
for mining of molecular dynamics results are presented also.

\end{abstract}

\end{frontmatter}

\section*{Introduction}

Is everything clear with the entropy growth? It seems that it is
not. A collection of problem statements and approaches was
published by Physica A on the eve of the millennium
\cite{Leb,Lieb,Pri}. Very recently, V.L. Ginzburg in his Nobel
Lecture characterized this problem as one of the greatest
challenges for physicists:
\begin{quotation} The ``great problems" are, first, the increase in
entropy, time irreversibility, and the ``time arrow"
\cite{Ginzburg}.
\end{quotation}
We usually describe the time arrow as disorder increase, and
measure disorder by the (logarithm of) phase volume following the
famous Boltzmann epitaph $$S=k \ln W$$ $W$ is the volume of an
ensemble, and $S$ is the entropy of this ensemble. The
ensemble--based point of view was expressed recently in the
following reasoning (\cite{GorKar}, p. 329):

\begin{quotation}
The well known question of what has more order,  a fine castle or a
pile of stones, has a profound answer:  It depends on which pile you
mean.  If ``piles'' are thought as all configurations of stones
which are not castles, then there are many more such piles, and so
there is less order in such a pile. However, if these are specially
and uniquely placed stones (for example, a garden of stones), then
there is the same amount of order in such a pile  as in a fine
castle. {\it Not a specific  configuration is important but an
assembly of configurations embraced by one notion.}
\end{quotation}

It seems to be true, but it is not the whole truth. In this paper
the ensemble--based point of view will be complemented by the
state--based one: The notions of order and disorder can describe
not only ensembles, but points also.

The following {\it toy--example} gives us a nice possibility to
understand the difference between the state--based and the
ensemble--based point of view, and helps us to learn how the
measure of order and disorder depends on the human activity and
perspective as well as on a state itself. Most of people are
familiar with the situation described in this example.

In book \cite{oet} the picture of ``order" after intensive play of
four children is presented to illustrate the idea: the definition
of order depends on a point of view, and the same set of positions
and orientations of toys may serve as a representative of rather
big ensemble of equivalent disorders (``parents--room"), or as an
almost unique configuration that changes sense after small change
(``children--room"). Children implicitly use the positions and
orientations of all their toys in their play. For parents, these
differences are not important. The same room (a state) produces
different ensembles, it depends on perspective. The notion
``order" distinguishes wide ensemble of the parents--room (big
volume, disorder) from narrow ensemble of the children--room
(small volume, order), and the entropy measures this difference.
This situation should be reflected on deeply before entering any
discussion about order--disorder measurement.

This difference between the parents--room and the children-room
can be formalized by the volumes of equivalent configurations. For
the parents--room, it seems to be larger, because the
parents--equivalence is coarser (they use ``other variables" for
description of the state of the room). Here we meet the important
operation that replaces a state (a point) by an ensemble. The
simplest formal version of this operation is the so-called
``fattening": in a metric space with metric $\rho(x,y)$ for any
set $A$ and $\varepsilon >0$ the $\varepsilon$-fattening of $A$ is
the set
\begin{equation}\label{fatt}
A_{\varepsilon}=\{x:\rho(x,y)<\varepsilon \; \mbox{for some} \; y
\in A \}.
\end{equation}
The set $A_{\varepsilon}$ includes all points that belong to $A$
``with accuracy ${\varepsilon}$."\footnote{The fattening is
similar to the Ehrenfest's coarse--graining
\cite{Ehrenfest,GKOeTPRE2001}.} Our first attempt to describe the
difference between the parents--room and the children--room is the
hypothesis that these ensembles are results of
$\varepsilon$-fattening for the same state (a point), but with
significantly different ${\varepsilon}$. The volume of the
parents--room--ensemble is much higher than the volume of the
children--room--ensemble.

This point of view is not the final one. Later, in this paper, it
will be complemented by the permutation analysis: the
parents--room has more permutation symmetry than the
children--room, and this causes significant difference between
their $\varepsilon$-fattening even for the same $\varepsilon$. The
symmetrization occurs to be the most important operation for
understanding of the difference between thermodynamic order and
disorder.

In this paper, the order-disorder problem is represented as a
problem of relation between distance and measure. The main focus
of our consideration is the effect of {\it order--disorder
separation}: for systems with a large number of particles the
available phase space (or configuration space) can be divided into
two parts. One part has microscopically small diameter (part $D$,
disorder), another part (part $O$, order) has microscopically
small measure (volume). We call a quantity {\it microscopically
small}, if it tends to 0 when the number of particles tends to
$\infty$. Of course, a proper normalization of the volume and
distance is assumed. As a consequence of the order--disorder
separation it is worth to mention the existence of  such a
microscopically small $\varepsilon>0$ that for each point $x$ from
the part $D$ its $\varepsilon$-fattening $\{x\}_{\varepsilon>0}$
includes almost all volume (the rest of the volume is
microscopically small).

We follow the idea of {\it thin--thick decomposition} (see M.
Gromov book \cite{Gromov99}, p. 124). The effect of
order--disorder separation is one of the {\it measure
concentration effects}. The geometry of spaces with finite, but
very large dimension has some interesting features that simplify
the asymptotic picture in comparison both with the small
dimensional, and the infinite-dimensional pictures. The typical
questions refer to various asymptotic relations between the
Lebesgue measure and the Euclidean distance. Recently, the effects
of this kind have been studied very intensively
\cite{Mil,Gromov99,GromWaist}. Some links between concentration of
measure and works of Boltzmann, Maxwell, Gibbs, and Ehrenfest are
presented below (nothing is absolutely new).

For the measure concentration, that leads to order--disorder
separation the permutation symmetry between particles (PI -- {\it
Permutation Invariance}) is important.

The paper has the following structure. In the next section, two
classical examples of measure concentration are presented: the
waist (or Maxwell) concentration of all the volume of
multidimensional spheres near equators, and the boundary (or
Gibbs) concentration of the volume of multidimensional balls near
boundaries (spheres). In Sec.~2, the Feynmann analysis of an
example of order increase is collated \cite{Feinman}. The
order--disorder separation for the Feynmann example is
demonstrated in Sec.~3.

Below, we discuss the statistical idea of order/disorder only. It is
based on the analysis of differences between less probable/more
probable events for large systems. There exist many other notions of
order/disorder, most important of them is the presence/absence of a
regular structure. We don't touch them in this paper.

\section{The classical measure concentration effects}\label{Sec1}

For large dimension $n$, the main part of the volume of the unit
$n$-dimensional ball $B^n$ is concentrated in a small neighborhood
of its boundary, that is the unit sphere $S^{n-1}$. This simple, but
very seminal fact can be demonstrated, as follows. Let us use the
normalized volume $|\bullet|$: $|B^n|=1$. The correspondent
(normalized) surface area of a unit sphere is a constant $C_n$,
$|B^n| = \int_0^1 C_nr^{n-1} \, \D r$, hence, $C_n=n$. The volume of
the part of $B^n$ inside the $\varepsilon$-neighborhood of $S^{n-1}$
is
\begin{equation}
V_{\varepsilon}=1-\int_0^{1-\varepsilon} n r^{n-1} \, \D r =
1-(1-\varepsilon)^n  .
\end{equation}
For small $\varepsilon$ and large $n$ (say, $n > 1/\varepsilon$)
we obtain the exponential estimate:
\begin{equation}\label{expest}
V_{\varepsilon}=1-(1-\varepsilon)^{ \frac{1}{\varepsilon}n
\varepsilon} \approx1- \exp(- n\varepsilon).
\end{equation}
It implies that for given $\varepsilon$ and $n\rightarrow \infty$
the volume $V_{\varepsilon} \rightarrow 1$ as  $1-\exp(-
n\varepsilon)$ (exponentially).

The sphere $S^{n-1}$ can be considered as the isoenergetic surface
for a very simple energy function, $E=\sum_{i=1}^n x_i^2$, that is,
for kinetic energy of $n$ classical particles on a line, or for
potential energy of $n$ simplest classical oscillators. Of course,
the observed concentration theorem could be proved for more general
energy functions.  Usually these generalizations are formulated as
theorems of ensemble equivalence: for large $n$ the canonical
ensemble (ensemble with probability distribution that maximizes the
entropy functional for a given average energy value) is equivalent
to the microcanonical ensemble (that is equidistribution on the
isoenergetic surface with respect to invariant Liouville measure).
A.I. Khinchin (1943) \cite{Khin} describes the probabilistic theory
of ensemble equivalence  when energy is a ``sum function", this
means that the system consists of a large number of noninteracting
subsystems. This type of concentration we call the Gibbs
concentration. The analysis of ensemble equivalence and
nonequivalence is presented in Ref.~\cite{Ellis04} with relevant
references.

The same type of reasoning can be applied to a hemisphere
$H^{n-1}=\{x \in S^{n-1}: x_1 \geq 0 \}$: for large $n$ almost all
measure of the hemisphere $H^n$ is concentrated near its boundary
$S^{n-2}=\{x \in S^{n-1}: x_1 = 0 \}$. Hence, almost all measure of
$S^{n-1}$ is concentrated near its $n-2$-dimensional equator. The
exponential estimate of the type (\ref{expest}) is also valid. The
well known application of this ``waist concentration" is the Maxwell
distribution for particle velocity: if the $n$-particle system in
the velocity space is equidistributed on the sphere of  radius
$R^2=\sum_{i=1}^n v_i^2=3nkT/m$, then, for large $n$, the velocity
of one particle will be distributed due to the Maxwell distribution.
The distribution of $v_1$ has the Gaussian density $\frac{1}{\sqrt{2
\pi} \sigma} \exp(-v_1^2/2 \sigma^2)$, where $\sigma^2=kT/m$. In
other term, the projection of uniform distribution from the unit
sphere $S^n$ onto the first axis has, for large $n$, the narrow
(almost) Gaussian distribution $\frac{1}{\sqrt{2 \pi} \sigma}
\exp(-v_1^2/2 \sigma^2)$, where $\sigma=1/\sqrt{n}$.

Due to ensemble equivalence this Maxwell concentration might be
demonstrated as concentration of the projection of the
equidistribution in the ball $B^n$ on a line. This projection is a
probability distribution on the segment $[-1,1]$ with the density
$\sim (\sqrt{1-x^2})^n$. For large $n$, $(\sqrt{1-x^2})^n \approx
\exp(-nx^2/2)$, and the projection density approaches the Gaussian
distribution $\sqrt{\frac{n}{2 \pi }}\exp(-nx^2/2)$ with the
standard deviation $\sigma=1/\sqrt{n}$.

The waist concentration holds not only for projection on
coordinate axis, but for any (nonlinear) Lipschitz function $F(x)$
with Lipschitz constant 1 ($|F(x)-F(y)|\leq |x-y|$): for large
$n$, the values of such a function on $S^n$ are concentrated in a
$\frac{1}{\sqrt{n}}$-small interval around the median value
$\overline{F}$ defined by the following statement: $$\PP(F(x)\geq
\overline{F})\geq \frac{1}{2} \; \mbox{and} \; \PP(F(x)\leq
\overline{F})\geq \frac{1}{2}.$$ It is the Levy theorem
\cite{Levy}.

The Maxwell distribution was known before statistical mechanics
was developed by Gibbs (and almost at the same time and
independently by Einstein). The waist concentration, in this
sense, was discovered by Maxwell.

Let us mention one important property of the waist concentration:
the points on the sphere are distributed uniformly, and are
equivalent in any reasonable sense: the measure is concentrated near
{\it every} equator. In one-dimensional projections (both linear and
general Lipschitz) this symmetry is destroyed, and there are
distinguished points, the median and its $\frac{1}{\sqrt{n}}$-small
neighborhood. The complement of this set has small measure, and this
set of distinguished points has small diameter; and, of course, a
small vicinity of any distinguished point has the same property, it
has the ``almost full" measure, and the small diameter. It does not
matter, if this projection is linear or not, only the Lipschitz
property is important. It makes no difference if the projection is
not one-dimensional: for any given dimension and for the number of
degrees of freedom $n\rightarrow \infty$ the result is the same. The
final results concerning the waist concentration for different
possible relations between $n$ and dimension of projection were
obtained by M. Gromov \cite{GromWaist}.

We can call the distinguished points as ``thermalized" states, or
``near-\-equili\-brium" states, but initially, on the
multidimensional sphere, all the states are equivalent, and the
distinguished points of measure concentration emerge only in a
macroscopic projection. In the following section we will present
the order--disorder separation for microscopic state.

\section{Strong order--disorder separation for symmetric microscopic states}

All the classical statistical physics is the theory of symmetric
ensembles: the density $\rho(x_1,x_2, \ldots x_n)$ of the full
multi-particle probability distribution is symmetric with respect
to particles permutations (here $x_i$ is a phase point for the
$i$-th particle). In this section, we demonstrate the
concentration effect that emerges in the projection of the phase
space (or configuration space) of $n$ particles onto the space of
permutations orbits. The $n$-particle space is $P^n$, where $P$ is
an one-particle space. The space of orbits can be presented as the
space of $n$-point subsets in the one-particle space $P$ (in the
measure and distance discussion for continuous spaces we can
neglect the degenerate case when positions of some particles
coincide).

\subsection{Feynman's blue and white atoms mixing}

Let us start from a simplest example of blue and white atoms
mixing analyzed in the book ``The Character of Physical Law," by
R. Feynman \cite{Feinman}.

\begin{quotation}
You have atoms of two different kinds (it's ridiculous, but let's
call them blue and white) jiggling all the time in thermal motion.
If we were to start from the beginning we should have mostly atoms
of one kind on one side, and atoms of other kind on the other side.
Now these atoms are jiggling around, billions and billions of them,
and if we start them with one kind all on one side, and the other
kind on the other side, we see that in their perpetual irregular
motions they will get mixed up, and that is why the water becomes
more or less uniformly blue. ...

If you start with a thing that is separated and make irregular
changes, it does get more uniform. But if it starts uniform and you
make irregular changes, it does not get separated. It {\it could }
get separated. It is not against the law of physics that the
molecules bounce around so that they separate. It is just unlikely.
It would never happen in a million years. And that is the answer.
\end{quotation}

This discussion is interesting not only by the clearly explained
thing, but by the carefully hidden things also. Let $P$ be the box
where the atoms move. The configuration space is $P^n$, where $n$
is the number of particles. The separated configurations (``with
one kind all on one side, and the other kind on the other side")
form an ensemble (a ``drop") with volume $2^n$ times smaller than
the whole volume of $P^n$. The concentration effects in the
velocity spaces allow us to represent the correspondent ensemble
in a phase space as a drop with a constant density inside it  also
(for example, with equidistribution in a velocity ball). It is
convenient for discussion. The volume of this drop is $2^n$ times
smaller than the equilibrium volume (hence, the density is $2^n$
times larger). This volume is conserved in the mechanical motion.
Hence, after some time this ensemble become more mixed, but
remains ``oil in water", that is, a phase space drop with the same
volume and density. In the sense of ensembles it is not a
``uniform" ensemble, and if somebody (the Maxwell demon, for
example) carefully  inverted all the velocities, this ensemble
would return to the initial separated state.

What does Feynmann mean: ``starting from homogeneous state we
never will get the separation... ?" It is absolutely new ensemble
``uniform states", it is not a result of the initial ensemble
evolution. Starting from the initial separated state we do reach
some of the ``uniform states", but not all such states. The phase
volume is different. For ``all uniform states" it is $2^n$ larger,
where $n$ is the number of particles. How can we get {\it all} the
uniform states (ensemble $U$) from the states we can reach from
our ordered states (ensemble $O$)?

And here Feynman uses an unexpected new notion, {\it irregular
changes}: ``If you start with a thing that is separated and make
irregular changes, it does get more uniform." And back: ``if it
starts uniform and you make irregular changes, it does not get
separated."

Who and how makes these  irregular changes and what does it mean?
The small portion of irregular changes makes the mixed ``oil in
water" ensemble strictly uniform. Where did this concept come
from? We can find a source of this idea in the coarse-graining.

The idea of coarse-graining dates back to P.\ and T.\ Ehrenfests,
and it has been most clearly expressed in their famous paper of
1911 \cite{Ehrenfest}. Ehrenfests considered a partition of the
phase space into small cells, and they have suggested to
supplement the motions of the phase space ensemble due to the
Liouville equation with ``shaking'' - averaging of the density of
the ensemble over the phase cells. As a result of this process,
the convergence to the equilibrium becomes uniform out of the
convergence in average. It is the fattening that we mentioned in
Introduction. This ``fattening-based" approach was developed into
a general technique of nonequilibrium thermodynamics
\cite{GKOeTPRE2001,GKIOeNONNEWT2001}. What is the physical nature
of the $\varepsilon$-fattening? First interpretation is noise, any
kind of small noise, small perturbations, and $\varepsilon$ is the
amplitude of this noise. Another interpretation of $\varepsilon$
is the possible accuracy of measurement and control.

But there is a purely mechanical effect: we start from the state
with small volume of its $\varepsilon$-fattening and after some
time of motion the system typically reaches states with large
volume of their $\varepsilon$-fattening.

After some time of mechanical motion a typical state (a point, not
an ensemble) becomes ``thick": permutation symmetrization with
microscopically small fattening transforms this point into an
uniform ensemble. The explanation of this effect is based on the
study of the geometry of a multidimensional simplex that we
perform in the next subsection.

Let us watch blue particles only, and an one-dimensional box
$P=[0,1]$ (in the direction of separation $x$). In order to
represent the set\footnote{Positions of some particles can
coincide, and, rigorously, a ``set" of $n$ particles forms an {\it
unordered tuple}. An unordered tuple of length $n$ of set $P$ is a
unordered selection with possible repetitions of set $P$ and is
represented by a sorted list of length $n$. In one-dimensional
case it is convenient to sort positions (numbers) in ascending
order.} of $n$ particles as a point in a standard simplex, we
introduce {\it symmetric coordinates} for $n$-particle systems.
Let us enumerate particles in the order of $x$ value: $0=x_0 \leq
x_1 \leq x_2 \leq \ldots \leq x_n \leq x_{n+1}=1$. Symmetric
coordinates are:
\begin{equation}\label{simplPresent}
s_i = x_{i}-x_{i-1},
\end{equation}
where $i=1, \ldots n+1$. Unordered $n$-particle states form in
coordinate $s_i$ a standard simplex $\Delta_n$: $s_i \geq 0$,
$\sum_i s_i =1$. The configuration volume transforms into a
uniform distribution in this simplex with a constant density $n!$.

Of course, it is possible to study the space of permutation orbits
as a quotient space endowed by quotient metrics. For the Euclidean
metric in the one-particle space, the quotient metrics is
\begin{equation}\label{quotient}
d_Q(\{x_1,x_2, \ldots ,x_n\},\{y_1,y_2, \ldots ,y_n\})=
\left[\min_{\sigma} \left\{\sum_{i=1}^n
\|x_i-y_{\sigma(i)}\|^2\right\}\right]^{1/2},
\end{equation}
where minimum is calculated for the set of all $n$-particle
permutations $\sigma$. Nevertheless, the use of symmetric
coordinates is more transparent. There are several other symmetric
representation of $n$-particle systems: measure representation and
functional (distance) representation. They are discussed below.

\subsection{Distance--measure relations in large--dimensional simplex}

Let us consider an $n$-dimensional standard simplex $\Delta_n$.
The normalized equidistribution in $\Delta_n$ has the constant
density $n!$. We call the correspondent probability measure the
{\it normalized volume}, and use notation $\mbox{Vol}(\bullet)$:
$\mbox{Vol}(\Delta_n)=1$. When discussing the probability, we
identify the probability of an event $\mathbf{P}\{\bullet\}$ with
the volume of a correspondent set $\mbox{Vol}(\bullet)$.

For large $n$, almost all volume of the simplex $\Delta_n$ is
concentrated in a small neighborhood of the center of $\Delta_n$,
near the point $c= \left({1 \over n}, {1 \over n}, \ldots, {1 \over
n}\right)$. The Euclidean radius of this neighborhood $R$ can be
chosen of order $\sim n^{-1/2}$. A projection of an $n$-dimensional
Euclidean ball with unit radius on a line is concentrated in an
interval of length $\sim n^{-1/2}$. It is the Maxwell (the waist)
concentration. Hence, any projection of an $n$-dimensional standard
simplex on a line is concentrated within an interval of length $\sim
n^{-1}$. This is true not only for orthogonal projections, but for
any Lipschitz   functions with Lipschitz constant 1 (1-Lipschitz
functions), as it is for Levy concentration. (See \cite{Gromov99},
p. 235.)

In order to demonstrate the main concentration properties of a
simplex, let us start with the moment evaluation. The moments give
this estimate of concentration radius in simplex, but only power
estimates of deviations are achievable on this way. Let us follow
Chebyshev's inequality for positive random variable $\xi$:
$\mathbf{P}\{ \xi \geq a \} \leq \mathbf{E}( \xi) /a$, where
$\mathbf{E}( \xi)$ is the expectation of $\xi$ (the average).

The distribution density for value $s$ of one coordinate $s_i$ in
$n$-dimensional standard simplex   is $p_1(s)=n(1-s)^{n-1}$, the
mutual density function for two coordinates, $s_1, s_2$ is
$p_2(s_1,s_2)=n(n-1)(1-s_1-s_2)^{n-2}$, for $k$ coordinates $s_1,
s_2, \ldots, s_k$ ($k<n$) the mutual density is $$p_k(s_1, s_2,
\ldots, s_k)=\frac{n!}{(n-k)!}\left(1-\sum_{i=1}^k s_i
\right)^{n-k}.$$

The first moments are: $\mathbf{E}(s)=1/(n+1)=1/n +o(1/n)$,
$\mathbf{E}(s^2)=2/[(n+1)(n+2)]=2/n^2 +o(1/n^2)$,
$$\mbox{Var}(s)=\mathbf{E}(s^2)-(\mathbf{E}(s))^2=\frac{n}{(n+1)^2(n+2)}=
\frac{1}{n^2}+o\left(\frac{1}{n^2}\right),$$ and for $k<n$,
$$\mathbf{E}(s^k)=\frac{1}{C_n^k}=\frac{(n-k)!k!}{n!}=\frac{k!}{n^k}+o\left(\frac{1}{n^k}\right)$$
(the last equality holds for  any given $k$ and $n \rightarrow
\infty$). For the first mixed moments we get
$\mathbf{E}(s_1s_2)=1/[(n+1)(n+2)]=1/n^2+o(1/n^2)$,
$$\mbox{Cov}(s_1,s_2)=\mathbf{E}(s_1s_2)-\mathbf{E}(s_1)\mathbf{E}(s_2)=
-\frac{1}{(n+1)^2(n+2)}=-\frac{1}{n^3}+o\left(\frac{1}{n^3}\right),$$
and for the correlation coefficient
$$\mbox{Cor}(s_1,s_2)=\frac{\mbox{Cov}(s_1,s_2)}{\sqrt{\mbox{Var}(s_1)\mbox{Var}(s_1)}}=-\frac{1}{n}.$$
It is worth to mention that $\mbox{Cov}(s_1,s_2)$ has order
$n^{-3}$, $\mbox{Var}(s)$ has order $n^{-2}$, hence, correlations
between coordinates decrease as $n^{-1}$ (coordinates become
independent for large $n$, and correlation decrease is a symptom
of this independence). It is easy to calculate moments of the
square of the Euclidean radius $R^2=\sum_{i=1}^{n+1}
(s_i-\mathbf{E}(s_i))^2$, for example
$$\mathbf{E}(R^2)=(n+1)\mbox{Var}(s)=\frac{n}{(n+1)(n+2)}=\frac{1}{n}+o\left(\frac{1}{n}\right),$$
and the Chebyshev's inequality gives the simplest estimate:
\begin{equation}
\mbox{Vol}\{x \in \Delta_n : R^2>\rho ^2\}\leq \frac{1}{\rho^2 n},
\end{equation}
up to the leading order in $n$.

With the higher moments of $R^2$ we can obtain estimates with the
higher powers of $1/n$, but already a simple geometrical
consideration gives exponential estimates. For any $i=1 \ldots n$,
the part of $\Delta_n$, where $s_i \geq \varepsilon$, has the
normalized volume
\begin{equation}\label{wingvol}
\mbox{Vol}(s \in \Delta_n : s_i \geq \varepsilon) =
(1-\varepsilon)^n \approx \exp(-\varepsilon n).
\end{equation}
Hence, the set $K_{\varepsilon} \subset \Delta_n$, where $s_i <
\varepsilon$ for all $i=1,\ldots , n+1$, has the normalized volume
\begin{equation}\label{allwings}
V_{\varepsilon} \geq (1-(n+1)(1-\varepsilon)^n) \approx 1- n
\exp(-\varepsilon n).
\end{equation}
For any point $x=(s_1, \ldots, s_{n+1}) \in K_{\varepsilon}$ the
following inequality holds: $R^2=\sum_{i=1}^n s_i^2 \leq
\sum_{i=1}^n \varepsilon s_i = \varepsilon$. Therefore, the
intersection of $\Delta_n$ and a Euclidean ball $B_{\rho}^{(n+1)}$
with the center $c$ includes the set $K_{\varepsilon} $, if
$\varepsilon  \leq \rho ^2$. Hence, for the normalized volume of
this intersection, $W_{\rho}$, the following inequalities hold:
\begin{eqnarray}\label{simCon1}
&&\mbox{Vol}\{x \in \Delta_n : R^2>\rho ^2\}=W_{\rho} \geq
V_{\rho^2} \nonumber \\ && \geq (1-(n+1)(1-\rho^2)^n) \approx 1-n
\exp (- \rho^2 n).
\end{eqnarray}

The estimate (\ref{simCon1}) implies that for any given positive
constant $a<1$ there exists a positive constant $b$ such that
$W_{b\ln n/\sqrt{n}}>a$ for all $n$. In other words, for any given
share $a$ of the simplex volume there exists such a constant $b>0$
that the Euclidean ball $B_{b \ln n/\sqrt{n}}^{(n+1)}$ with the
center $c$ includes this part of the volume for all $n$. We can
guarantee with (\ref{simCon1}) that the radius of such a ball goes
to zero as $\ln n/\sqrt{n}$.

A precise analysis of the  concentration effects in $L_p$ balls
and in a standard symplex was performed in
\cite{Gromov99,Shlechtman}.

The concentration of a simplex measure in a small vicinity of its
center can be considered as an effect that is opposite  to the Gibbs
concentration of volume of a $n$-dimensional ball $B_n$ in a small
vicinity of its boundary, the sphere. On the other hand, it is
similar to the waist concentration. And now not only the values of
macroscopic projections can be separated onto two sets: one with a
microscopically small diameter, the other with a microscopically
small measure, but also the set of the symmetrized microscopic
states. The symmetrization with respect to particles permutations
plays the same role as the macroscopic projection. We can say now
that this {\it symmetrization is the main step in the micro--macro
transformation}.

\subsection{Symmetric coordinates for multidimensional phase
space} \label{SymND}

In order to demonstrate the same effect for one-particle
configuration space (or phase space) of non-unit dimension, let us
consider a product of $m$ simplices $\Delta_n$ for $m\sim
n^{\alpha}$ and some power ${\alpha}$. Euclidean diameter of
$\Delta_n^m$ grows with $m$ as $\sqrt{m} \sim n^{\alpha /2}$,
Euclidean diameter of the product of Euclidean balls
$(B_{\rho}^{(n+1)})^m$ is $2 \sqrt{m} \rho \sim \rho n^{\alpha
/2}$. For the normalized volume of the intersection $\Delta_n^m
\cap B_{\sqrt{m} \rho}^{m(n+1)}$ the following estimate holds:
\begin{eqnarray}
&&\mbox{Vol}\left(\Delta_n^m \cap B_{\sqrt{m}
\rho}^{m(n+1)}\right) \geq
 \mbox{Vol}\left(\Delta_n^m \cap
(B_{\rho}^{(n+1)})^m\right) \nonumber \\ &&\approx (1-n \exp (-
\rho \sqrt{n}))^m \sim 1- n^{1+\alpha}\exp (- \rho \sqrt{n}).
\end{eqnarray}
From this estimate it follows  that the strong order-disorder
separation holds for these Cartesian degrees of simplex also (if
$m\sim n^{\alpha}$): almost all volume belongs to an Euclidean
ball with the relatively small diameter $R \sim \rho n^{\alpha
/2}$. In order to include in this ball any given share of volume
we can choose $\rho \sim n^{-1/2}$ with appropriate value of the
prefactor. Therefore, the correspondent relation of diameters
$R/$Diam($\Delta_n^m$) goes to zero as $n^{-1/2}$.

Let one-particle space be $k$-dimensional unit cube $Q_k$. The space
for $n$-particle system is $(Q_k)^n$. We produce the symmetric map
of $(Q_k)^n$ onto product of $n^{1/k}$ dimensional simplices
$(\Delta_{n^{1/k}})^{kn^{(n-1)/k}}$ as follows. Let $\xi_i$, $i=1,
\ldots k$, $0\leq \xi \leq 1$ be coordinates in $Q_k$. With each
coordinate axis we construct a projection of $(Q_k)^n$ onto
$(\Delta_{n^{1/k}})^{n^{(n-1)/k}}$. The product of $k$ such
projections is the resulting map $(Q_k)^n \rightarrow
(\Delta_{n^{1/k}})^{kn^{(n-1)/k}}$.

For $\xi_k$, this projection is the top floor of the ``staged
tower" of symmetric coordinates. Let us first enumerate particle
in the order of $\xi_1$ value: $0=x_0 \leq x_1 \leq x_2 \leq
\ldots \leq x_n \leq x_{n+1}=1$. First set (the ground floor of
the ``staged tower") of symmetric coordinates is: $s_i =
x_{i}-x_{i-1}$, where $i=1, \ldots n+1$.

Let us divide the particles into $n^{1/k}$ groups $G^l$, $l=1,
\ldots, n^{1/k}$ with $n^{(k-1)/k}$ elements in each group in the
same order: first $n^{(k-1)/k}$ particles with coordinates $\xi_1
= x_1, x_2, \ldots x_{n^{(k-1)/k}}$ belong to the first group,
$G^1$, then follow $n^{(k-1)/k}$ particles from the second group,
etc. Let us enumerate particle of each group in the order of
$\xi_2$ value: $0=x^l_0 \leq x^l_1 \leq x^l_2 \leq \ldots \leq
x^l_{n^{(k-1)/k}} \leq x_{{n^{(k-1)/k}}+1}=1$, where superscript
$l$ is the group number.\footnote{Of course, it is more rigorous
to speak about integer parts of numbers: $G^1$ consists of
$\mbox{IntegerPart}(n^{(k-1)/k})$ elements,  $G^2$ consists of
$\mbox{IntegerPart}(2n^{(k-1)/k})-\mbox{IntegerPart}(n^{(k-1)/k})$
elements, and so on, but it adds nothing to the sense, only the
notations become cumbersome.}

The first floor of the ``staged tower" consists of $n^{1/k}$ sets
of symmetric coordinates $s^l_i=x^l_{i}-x^l_{i-1}$. After that, we
can divide each $G^l$ into $n^{1/k}$ groups  $G^{lm}$, $m=1,
\ldots, n^{1/k}$ with $n^{(k-2)/k}$ elements in each group in the
order of $\xi_2$ value. Let us enumerate particles of each group
in the order of $\xi_3$ value: $0=x^{lm}_0 \leq x^{lm}_1 \leq
x^{lm}_2 \leq \ldots \leq x^{lm}_{n^{(k-2)/k}} \leq
x^{lm}_{{n^{(k-2)/k}}+1}=1$. The second floor consists of
$n^{2/k}$ sets of symmetric coordinates
$s^{lm}_i=x^{lm}_{i}-x^{lm}_{i-1}$. Finally, we get $k$ floors
(from the ground to $(k-1)$st one). The floor number $j$ ($j=0,
\ldots k-1$) consists of $n^{j/k}$ groups of symmetric coordinates
with $n^{(k-j)/k}$ coordinates in each group. These coordinates
are non-negative, their sums in groups are equal to 1. Therefore,
each floor represents a $n$-dimensional polyhedron that is a
product of $n^{j/k}$ standard simplices $\Delta_{n^{(k-j)/k}}$ of
dimension $n^{(k-j)/k}$, and the whole tower represents the
following product of simplices
\begin{equation}
\Omega_{k,n} =\prod_{j=0}^{k-1} \Omega_j, \; \mbox{where} \;
\Omega_j=(\Delta_{n^{(k-j)/k}})^{n^{j/k}}.
\end{equation}
We are interested in the $(k-1)$st floor that corresponds to
$\xi_k$. It is $\Omega_{k-1}=(\Delta_{n^{1/k}})^{n^{(n-1)/k}}$.
Analogous projection for other $\xi_k$ could be obtained by
coordinates permutation (cyclic).

We see that for a $k$-dimensional one-particle space the result is
qualitatively the same as for one-dimensional. The only difference
is that here the estimates guarantee that the relative Euclidean
radius (that is, the relation of the radius to the diameter of the
whole space) of the set, where an arbitrary part $a<1$ of measure
is concentrated, tends to zero as $1/\sqrt{d}$, instead of
$1/\sqrt{n}$. Here $d$ is the dimension of one simplex from the
product, that is, $d=n^{1/k}$ and the relative radius goes to zero
as $n^{-1/(2k)}$.

This change of order reflects a simple fact: the typical distance
from a particle to the nearest particles in dimension $k$ is $\sim
n^{-1/k}$. After summation of $d$ squares of such variables we get
the square of radius: $R^2 \sim n^{-1/k}$. Then we take the
$n^{(k-1)/k}$th power of $d$-dimensional simplex and of the ball
from this simplex also. The relation of the Euclidean radii does
not change after this operation. It remains $\sim n^{-1/(2k)}$.

The same results hold for one-particle space $P$ that is not a
cube, but a bi-Lipschitz image of a cube, or can be covered by
finite number of such images. We discuss much more general
metric-measure ($mm$) spaces in the next subsection.

\subsection{Other natural distances on symmetrized states}

A metric space $P$ with distance $d(x,y)$ and a given measure
$\mu$ on $P$ is a $mm$-space \cite{Gromov99}, if every metric ball
is measurable. In this section we discuss distribution of
particles in a $mm$-space $P$ with a probability measure $\mu$,
hence, $\mu(P)=1$.

We assume that $P$ is compact\footnote{Generalization of most
statements to complete, but non-compact $mm$-spaces (for example,
to important case of locally compact space) is often possible
because the probability measure $\mu$ is concentrated on a compact
subset of $P$ up to any given accuracy, and after cutting a
``tail" of distribution $\mu$ we can return to compact space. The
theory of large deviations and equidistribution in general spaces
is presented in Refs. \cite{Ellis85,Ellis95}.}
 and, hence, has a finite
diameter. The space of (Radon) measures on $P$ is $C^*(P)$, that
is the conjugated space to the space of continuous functions
$C(P)$ on $P$. The action of a measure $\nu \in C^*(P)$ on a
function $f \in C(P)$ is the number $[\nu,f]$. The action of
probability measure $\mu$ on $f$ is the expectation: $[\mu,f]=
\mathbf{E}(f)$. The probability measures on $P$ are positive and
normalized elements of $C^*(P)$. For measures we use the weak$^*$
convergence: $\mu_i \rightarrow \mu_0$ if $[\mu_i,f] \rightarrow
[\mu_0,f]$ for every continuous function $f$.

An unordered tuple of $n$ points (``particles") in a $mm$-space
$P$ can be represented as a probability measure:
\begin{equation}\label{measrepr}
\{x_1,x_2, \ldots , x_n\} \mapsto \mu_{x_1,x_2, \ldots , x_n} =
\frac{1}{n}\sum_{i=1}^n \delta_{x_i},
\end{equation}
 where $\delta_{x_i}$ is a
unit measure concentrated at the point $x_i$ ($\delta$-function).
The {\it law of large numbers} states that $\mu_{x_1,x_2, \ldots ,
x_n} \rightarrow \mu $ for almost all sequences $\{x_1,x_2, \ldots
, x_n, \ldots \} \in P^{\infty}$. ``Almost all" means here: the
set of exceptions has zero measure. Let $f$ be an arbitrary
bounded continuous function on $P$. The standard law of large
numbers for a random variable $f$ and the probability space $P$
immediately gives the weak$^*$ convergence: the sequence of
averages $\langle f \rangle _n=\frac{1}{n}\sum_{i=1}^n f(x_i)$
converges to the average value of $f$ with respect to the
probability measure $\mu$, that is, to the expectation
$\mathbf{E}(f)$.

For each domain $W \subset P$ with non-zero measure $\mu (W)$ the
probability that all particles are outside $W$ is
\begin{equation}\label{generalwing}
\mathbf{P}\{x_i \notin W : i=1, \ldots n \}= (1-\mu (W))^n .
\end{equation}
 This
estimate is analogous to the estimate of the volume of one wing of
the $n$-dimensional standard simplex (\ref{wingvol}).

Moreover, {\it all} balls $B_{\rho}$ of given radius $\rho$ are
nonempty with almost unit probability. Let us take a $\rho/2$ net
$\{y_1, \ldots, y_m \}$ in $P$, where $m={\rm Cap}_{\rho/2}(P)$ is
the minimum number of points in $\rho/2$ net in $P$.  Each ball
$B_{\rho}$ in $P$ includes a ball $B_{\rho /2}(y_i)$ of radius $\rho
/ 2$ and center in one of the points $y_1, \ldots, y_m$. Therefore,
if there is a ball $B_{\rho} \subset P$ free of particles, then at
least one of the balls $B_{\rho /2}(y_i)$ ($i=1, \ldots, m$) is also
free of particles.
\begin{eqnarray}\label{estimBalls}
&&\mathbf{P}\{\mbox{Each ball} \; B_{\rho} \subset P \;
\mbox{includes a particle} \} \nonumber \\ && \geq 1 -
\sum_{i=1}^{{\rm Cap}_{\rho/2}(P)} (1- \mu(B_{\rho /2}(y_i))^n
\geq 1 - {\rm Cap}_{\rho/2}(P)(1 - \underline{\mu}(\rho / 2))^n,
\end{eqnarray}
where $\underline{\mu}(\rho)= \inf \{\mu(B_{\rho})(y) : y \in P \}$.
This estimate is similar to the estimate (\ref{allwings}) of the
joint volume of the $n$-dimensional simplex wings. In the final
estimate (\ref{estimBalls}) two characteristics of the $mm$-space
$P$ are used: the minimum number of points in $\rho/2$ net, ${\rm
Cap}_{\rho/2}(P)$, and the minimal volume of a ball of radius $\rho
/2$, $\underline{\mu}(\rho / 2)$. There is also a difference between
(\ref{allwings}) and (\ref{estimBalls}): the analogue for the
``number of wings" for the last estimate, ${\rm Cap}_{\rho/2}(P)$,
does not depend on $n$.

Natural metrization of the space of probability measures on $P$ in
the weak$^*$ convergence gives the following metric
\cite{Shir,Gromov99}:
\begin{equation}\label{measureSDIST}
Lid(\nu,\eta)=\sup _f|[\nu - \eta,f]|,
\end{equation}
where $f$ runs over all 1-Lipschitz functions on $P$.

Another, functional representation of an unordered tuple of $n$
points (``particles") in a $mm$-space $P$ is a continuous function
\begin{equation}\label{funcrepr}
\{x_1,x_2, \ldots , x_n\} \mapsto f_{x_1,x_2, \ldots , x_n}:
f_{x_1,x_2, \ldots , x_n}(x)=\min_{i=1, \ldots, n} d(x,x_i).
\end{equation}
This functional representation is an exact analogue for the
simplex representation (\ref{simplPresent}): in one-dimensional
case the maximum norm of the function $f_{x_1,x_2, \ldots , x_n}$
is $\frac{1}{2}\max_i s_i$, the average of $|f_{x_1,x_2, \ldots ,
x_n}(x)|^p$ is $\frac{1}{2^p(p+1)}\sum_i s_i^{p+1}$. Particularly,
the square of the Euclidean (i.e. $L_2$) norm in a simplex is
proportional to $L_1$ norm of the function $f_{x_1,x_2, \ldots ,
x_n}$.

For this representation the estimate (\ref{estimBalls}) has a
simple form
\begin{eqnarray}\label{estimFunk}
&&\mathbf{P}\{\|f_{x_1,x_2, \ldots , x_n} \|_{L_{\infty}} < \rho
\} \geq 1 - \sum_{i=1}^m (1- \mu(B_{\rho /2}(y_i))^n \nonumber \\
&& \geq 1 - m(1 - \underline{\mu}(\rho / 2))^n,
\end{eqnarray}
where $\|\bullet\|_{L_{\infty}}$ is the maximum norm.

In many practically important $mm$-spaces $P$ the volume of balls
is of order  $\rho^k$ for some power $k>0$: $\inf
\mu(B_{\rho})/\rho^k =a>0$. In that case, for sufficiently small
$\rho$ and large $n$ $$(1 - \underline{\mu}(\rho / 2))^n \approx
\exp(-an(\rho /2)^k)$$ and

\begin{equation}\label{estimFunk1}
\mathbf{P}\{\|f_{x_1,x_2, \ldots , x_n} \|_{L_{\infty}} < \rho \}
\geq 1-{\rm Cap}_{\rho/2}(P) \exp(-an(\rho /2)^k).
\end{equation}
where  ${\rm Cap}_{\rho/2}(P)$ is the minimum number of points in
$\rho/2$ net in $P$.

For $L_1$ norm of $f_{x_1,x_2, \ldots , x_n}$ (analogue for the
square of the Euclidean norm in a simplex of symmetrical
coordinates) we obtain the estimates $$\|f_{x_1,x_2, \ldots , x_n}
\|_{L_1} = \mathbf{E}(f_{x_1,x_2, \ldots , x_n}) \leq \max_P
|f_{x_1,x_2, \ldots , x_n}(x)|$$ and
\begin{eqnarray}
&&\mathbf{P}\{\|f_{x_1,x_2, \ldots , x_n} \|_{L_1} \leq b \} \geq
\mathbf{P}\{\|f_{x_1,x_2, \ldots , x_n} \|_{L_{\infty}} < b \}
\nonumber
\\ && \geq 1-{\rm Cap}_{b/2}(P)\exp(-an(b/2)^k).
\end{eqnarray}
And again we observe the simplex--type strong order--disorder
separation. In the maximum norm, the whole set of functions that
represent $n$--point tuples has diameter Diam$(P)$. The measure is
concentrated in a ball or radius  $\sim n^{-1/k}$
(\ref{estimFunk1}) (in the maximum norm also).

There exists a simple connection between measure (\ref{measrepr})
and functional  representations (\ref{funcrepr}) of $n$-particle
systems. Let the radius of a ball with the centre $x$ and volume
$\delta$ is continuous function $r_{\delta}(x)$ for any $\delta >
0$. For each continuous function $f(x)$ the $L_1$ function
$f^{\delta}(x)$ is defined:
\begin{equation}
f^{\delta}(x)=\left\{
\begin{array}{ll}
\frac{1}{\delta}, \: &\mbox{if} \: f(x)<r_{\delta}(x); \\ 0, & \:
\mbox{if} \: f(x) \geq r_{\delta}(x).
\end{array}\right.
\end{equation}
The distribution  $\frac{1}{n}f^{\delta}_{x_1,x_2, \ldots ,
x_n}\mu$ approximates the measure $\mu_{x_1,x_2, \ldots , x_n}$
(\ref{measrepr}) when $\delta \rightarrow 0$.

\subsection{Statistics of local structures}

In this paper, we discuss the statistics of large sets of particles
with permutation invariance. For spaces of $k$-particle sets, two
embeddings are considered: into space of measures (\ref{measrepr})
and into spaces of functions (\ref{funcrepr}). These embeddings are
useful for theoretical purposes, but for practical needs embeddings
into finite--dimension Euclidean space are necessary, as well as
systems of internal coordinate charts on spaces of $k$-particle sets
with one labeled point. In this subsection, we discuss the embedding
and coordinate choice and give a non-technical introduction into
statistical analysis on non-Euclidean metric spaces.

Molecular dynamics gives us many examples of particle
configurations. We never had such detailed information before, and
the question is how to process it with maximally useful output.
The classical approach of statistical physics is based on
$k$-particle distribution functions for small $k$. It is not
sufficient, for example, for the following problems.

Let the configuration of $n$ particles be given: we know all the
positions of molecules. For each particle (point $x$) and any
$k<n$ we define a $k$-particle local configuration, or {\it germ}
($k$-germ) of the configuration at $x$, that is the set of $k$
particles nearest to $x$ represented in the reference system with
origin $x$. The set of $k$-germs for all possible central
particles form a cloud of points in the space of $k$-germs. Are
there clusters or clots in this cloud? Is the distribution of
$k$-germs in physical space ($R^3$) homogeneous? If it is
heterogeneous, then how can we find boundaries between locally
homogeneous clusters?

For systems in isotropic conditions, instead of $k$-germs it is
necessary to consider orbits of $k$-germs under the action of
rotation group. Molecules with internal structure can also be
considered without principal problems (but with some technical
complications).

The problem of local heterogeneities in water is most attractive
\cite{HeterDeBene,HeterStanley1,HeterStanley2}. But even for hard
spheres systems the cluster boundaries localization is not
trivial.

It is not obvious, how many particles in local configuration
should we take into account: where the heterogeneities are hidden.
It is necessary to study statistics of $k$-germs for different $k$
and evaluate the informativity of transition from $k$ to $k+1$.

The classical {\it statistical geometry} gives some tools for
quantitative analysis of configuration structure. Important
sources of ideas and methods for the local configuration analysis
are the theory of random packing \cite{RANDOMPACK}, the molecular
geometry of liquids \cite{RANDOMPACK1} and the theory of
liquid-glass transition \cite{LiqGlass}. The main tool of
statistical geometry that is in wide use for molecular dynamics
data mining is the analysis of the Voronoi polyhedra and the
Delaunay simplices statistics \cite{VorDel1,VorDel2,VorDel3}. The
Voronoi polyhedron is the domain around a particle, such that all
points of this domain are closer to this particle than to any
other. A group of four particles, whose Voronoi polyhedra meet at
one vertex, forms another basic object of statistical geometry,
the Delaunay simplex. Statistics of the Voronoi polyhedra and the
Delaunay simplicec gives us information about local order in the
{\it nearest vicinity} of particles: the Voronoi polyhedron
describes the coordination of the nearest atomic environment while
the Delaunay simplex describes the shape of the cavities between
the nearest atoms.

In order to extend this vicinity to an arbitrary number of
neighbors and coordination spheres, we need the statistics of
$k$-germs. We propose systematic study of statistics of $k$-germs:
(nonlinear) principal component analysis, cluster analysis, and
analysis of the fields of obtained statistical characteristics in
the physical space-time. Many old question could be revisited in
this way, especially the problems of local heterogeneity.

The crucial question is the choice of a space where the
statistical analysis will be performed.

For statistical computations, the embedding of the space of germs
into Euclidean space is convenient. For any sequence of functions
in $R^m$, $\mathcal{F}=\{f_1, \ldots f_M\}$ let us define
\begin{equation}\label{coordgerm}
\mathcal{F}(x_1, \ldots x_k)=\left\{ \frac{1}{k} \sum_{i=1}^k
f_j(x_i)\right\}_{j=1}^M
\end{equation}

These coordinates serve for computation of distance $\rho$ between
germs (just a standard Euclidean distance in these coordinates can
be chosen; the second choice is the locally Euclidean Riemannian
metric, the geodesic distance).

For systems with rotational symmetry, it is necessary to study
statistics of rotational orbits of germs. The space of functions
spanned by $\{f_1, \ldots f_M\}$ should be rotationally invariant
and represented as a sum of irreducible subspaces. In this case, the
coordinate tuple  for a germ (\ref{coordgerm}) is a direct sum of
irreducible tensors, and it is easy to write the complete system of
rotational invariants and to define invariant distance on the space
of germs.

For this purpose, it is convenient to choose $\{f_1, \ldots f_M\}$
as eigenfunctions of a Schr\"odinger operator with central force,
sorted by eigenvalues and momentum (isotropic oscillator
eigenfunctions, for example). These eigenfunctions are spherical
harmonics multiplied on radial functions $f_j(\rho)$ that decay
when $\rho \to \infty$, hence, a far particle has less influence
on the distance between germs than the nearest one.

Statistics in Euclidean spaces with coordinates (\ref{coordgerm})
is not statistics of germs: the average of germs for this
Euclidean statistics is already not a germ. Let us consider the
space of germs as a non-Euclidean metric space with metric $\rho$.

Following Frechet \cite{Freshe}, we can define an average point
$\langle z \rangle$ for a finite subset $\{z_1, \ldots z_q \}$ of a
metric space $\mathcal{K}$ as a minimizer of average squared
distance
\begin{equation}\label{metrav}
\langle z \rangle = {\rm argmin}_{z \in \mathcal{K}}
\left\{\frac{1}{q} \sum_j \rho^2(z,z_j) \right\}.
\end{equation}
On the base of this approach, statistics on Riemannian spaces is
developed, from simple averaging to moments calculation and
definition of normal distribution \cite{Pennec}. For shape
statistics, the method of principal geodesic analysis is proposed,
that is a generalization of principal component analysis to the
manifold setting \cite{FletchShapPCA}.

We can interpret the Frechet averaging (\ref{metrav}) as
minimization of elastic energy of springs that connect data points
with an average point. The statistical analysis on metric spaces
may be represented as minimization of ``elastic energy"
\cite{nlprinc1,ZinovyevBook00,nlprinc2}. This energetic metaphor
works successfully  for model reduction problems, cluster analysis
and analysis of data with complex topology. Let us give a sketch
of this approach following \cite{TopGram}. For simplicity, we
consider a metric space embedded into Euclidean space with
Euclidean distance between points.

Let $G$ be a simple undirected graph with set of vertices $Y$ and
set of edges $E$. For $k \geq 2$ a $k$-star in $G$ is a subgraph
with $k+1$ vertices $y_{0,1, \ldots k} \in Y$ and $k$ edges
$\{(y_0, y_i) \ | \ i=1,\ldots k\} \subset E$. Suppose for each
$k\geq 2$, a family $S_k$ of $k$-stars in $G$ has been selected.
We call a graph $G$ with selected families of $k$-stars $S_k$ an
{\it elastic graph} if, for all $E^{(i)} \in E $ and $S^{(j)}_k
\in S_k$, the correspondent elasticity moduli $\lambda_i > 0$ and
$\mu_{kj}> 0$ are defined. Let  $E^{(i)}(0),E^{(i)}(1)$ be
vertices of an edge $E^{(i)}$ and $S^{(j)}_k (0),\ldots S^{(j)}_k
(k)$ be vertices of a $k$-star  $S^{(j)}_k $ (among them,
$S^{(j)}_k (0)$ is a central vertex).
 For any map $\phi:Y \to R^m$ the {\it energy of the
graph} is defined as
\begin{eqnarray}
U^{\phi}{(G)}&:=&\sum_{E^{(i)}} \lambda_i
\left\|\phi(E^{(i)}(0))-\phi(E^{(i)}(1)) \right\| ^2 \\ &&+
\sum_{S^{(j)}_k}\mu_{kj} \left\|\sum _ {i=1}^k \phi(S^{(j)}_k
(i))-k\phi(S^{(j)}_k (0)) \right\|^2. \nonumber
\end{eqnarray}

Very recently, a simple but important fact was noticed
\cite{Gusev04}: every system of elastic finite elements could be
represented by a system of springs, if we allow some springs to
have negative elasticity coefficients.  The energy of a $k$-star
$s_k$ in $R^m$ with $y_0$ in the centre and $k$ endpoints
$y_{1,\ldots k}$ is $u_{s_k}= \mu_{s_k}(\sum_{i=1}^k y_i - k
y_0)^2$, or, in the spring representation, $u_{s_k}=k\mu_{s_k}
\sum_{i=1}^k (y_i - y_0)^2 - \mu_{s_k} \sum_{i
> j} (y_i-y_j)^2$. Here we have $k$ positive springs with
coefficients $k\mu_{s_k}$ and $k(k-1)/1$ negative springs with
coefficients $-\mu_{s_k}$.

For a given map $\phi: Y \to R^m$ we divide the dataset $D$ into
subsets $K^y, \, y\in Y$. The set $K^y$ contains the data points
for which the node $\phi(y)$ is the closest one in $\phi(Y)$. The
{\it energy of approximation} is:
\begin{equation}
U^{\phi}_A(G,D):= \sum_{y \in Y} \sum_{ x \in K^y} w(x) \|x-
\phi(y)\|^2,
\end{equation}
where $w(x) \geq 0$ are the point weights.

The simple and very popular algorithm for minimisation of the energy
$U^{\phi}=U^{\phi}_A(G,D)+U^{\phi}{(G)}$ is the splitting algorithm,
in the spirit of the classical $k$-means clustering: for a given
system of sets $\{K^y \ | \ y \in Y \}$ we minimise $U^{\phi}$, then
for a given $\phi$ we find new $\{K^y\}$, and so on; stop when no
change occurs. This is the constrained minimisation: the nodes move
along the $k$-germs space embedded into Euclidean space, while the
distance in this example is Euclidean one. This algorithm gives a
local minimum, and the global minimisation problem arises. There are
many methods for improving the situation, but without guarantee of
the global minimisation.

The next problem is the elastic graph construction. Here we should
find a compromise between simplicity of graph topology, simplicity
of geometrical form for a given topology, and accuracy of
approximation. Geometrical complexity is measured by the graph
energy $U^{\phi}{(G)}$, and the error of approximation is measured
by the energy of approximation $U^{\phi}_A(G,D)$. Both are
included in the energy $U^{\phi}$. Topological complexity will be
represented by means of elementary transformations: it is the
length of the energetically optimal chain of elementary
transformation from a given set applied to initial simple graph.

Graph grammars \cite{NaglGram,Loewe} provide a well-developed
formalism for the description of elementary transformations. An
elastic graph grammar is presented as a set of production (or
substitution) rules. Each rule has a form $A \to B$, where $A$ and
$B$ are elastic graphs. When this rule is applied to an elastic
graph, a copy of $A$ is removed from the graph together with all its
incident edges and is replaced with a copy of $B$ with edges that
connect $B$ to graph. For a full description of this language we
need the notion of a {\it labeled graph}. Labels are necessary to
provide the proper connection between $B$ and the graph.

A link in the energetically optimal transformation chain is
constructing by finding a transformation application that gives the
largest energy descent (after an optimization step), then the next
link, and so on, until we achieve the desirable accuracy of
approximation, or the limit number of transformations (some other
termination criteria are also possible).

As a simple (but already rather powerful) example  we use a system
of two transformations: ``add a node to a node" and ``bisect an
edge." These transformations act on a class of {\it primitive
elastic graphs}:  all non-terminal nodes with $k$ edges are centers
of elastic k-stars, which form all the $k$-stars of the graph. For a
primitive elastic graph, the number of stars is equal to the number
of non-terminal nodes -- the graph topology prescribes the elastic
structure.

The transformation {\it ``add a node"} can be applied to any vertex
$y$ of $G$:  add a new node $z$ and a new edge $(y,z)$. The
transformation {\it ``bisect an edge"} is applicable to any pair of
graph vertices $y,y'$ connected by an edge $(y,y')$: delete edge
$(y,y')$, add a vertex $z$ and two edges, $(y,z)$ and $(z,y')$. The
transformation of elastic structure (change in the star list) is
induced by the change of topology, because the elastic graph is
primitive. This two--transformation grammar with energy minimization
builds {\it principal trees} (and principal curves, as a particular
case) for datasets.

For applications, it is useful to associate with these principal
trees one-dimensional continuums. Such a continuum consists of node
images $\phi(y)$ and of pieces of lines that connect images of
linked nodes.

The {\it first task} of $k$-germs statistical analysis is dimension
reduction. The method of choice here is principal component analysis
(PCA). Its linear version is now classics and textbook material
\cite{princ}, and nonlinear PCA is developed recently
\cite{HastieStuetzle89,nlprinc1,Kegl00,nlprinc2}. The methods of
elastic manifolds and graphs \cite{nlprinc2} does not require
Euclidean space of data. The {\it second task} is cluster analysis.
The described method of elastic graphs is a tool for nonlinear PCA,
and for cluster analysis, both.

The {\it third task} that is specific for statistical physics is
the analysis of $k$-germs distribution in physical space-time.
After that, we can discuss structural non-uniformity,
quasi-chemical representation of kinetics \cite{GorKar}, and many
other topics. Of course, additional topological information about
various bonds between particles could be added to this metric
description.

Internal coordinates on the space of germs are necessary for
gradient optimization of energy. Topologically, the space of
$k$-germs near a point $x$ can be defined as the space of
permutation orbits. Let us enumerate $k$ particles $x_1, x_2,
\ldots x_k$ nearest to the point $x$ in order of their distance to
$x$, $\rho_i=\|x_i-x\|$: $\rho_1<\rho_2< \ldots < \rho_k$. If all
particles are in generic positions, then any two distances are
distinct. This ordered representation $\{x_1, \ldots x_k \}$ has
discontinuity points when some $\rho_i$ coincide.

The following internal coordinates on the space of rotational
orbits of germs give generically a representation of these orbits
with discontinuity points when some $\rho_i$ coincide.

Let us enumerate $k$ particles $x_1, x_2, \ldots x_k$ nearest to the
point $x$ in order of their distance to $x$, $\rho_i=\|x_i-x\|$:
$\rho_1<\rho_2< \ldots < \rho_k$. We assume that all particles are
in generic positions, hence, any two distances are distinct and
three particles could not belong to one straight line. The distances
from $x_i$ to $x,x_1,x_2$ will be the main coordinates of the
$k$-germ. These are $3k-3$ numbers: $\{ \rho_i \}_{i=1,\ldots k}, \,
\{\rho'_j\}_{j=2,\ldots k}, \, \{ \rho''_l\}_{l=3,\ldots k}$, where
$\rho'_j=\|x_j-x_1\|, \, \rho''_l=\|x_l-x_2\|$. An additional set of
coordinates consists of $k-2$ signs, $\sigma_i=\pm 1, \, i=3, \ldots
k$. The triangle $\{x,x_1,x_2\}$ belongs to a plane $\Gamma$. This
plane divides the space into two half-spaces, $L_+$ and $L_-$. We
define the signs subscripts $L_{+,-}$ by triangle orientation
$\{x,x_1,x_2\}$ according to the standard ``screw rule" (or the
``right--hand rule"). The sign $\sigma_i=+1$, if $x_i \in L_+$,
$\sigma_i=-1$, if $x_i \in L_-$. Generically there are no particles
on $\Gamma$. The whole set of coordinates consists of $3k-3$ real
numbers and $k-2$ signs. If we, in addition to rotation symmetry,
assume the reflection symmetry, then there are only $k-3$ signs:
$\sigma'_j=\sigma_3 \sigma_j, \, j=4,\ldots k$. The worst violations
of the continuity condition for the proposed coordinates are jumps
of basis triangle $x,x_1,x_2$  near some of configurations. For
example, for a body--centered cubic lattice there are three
non-equivalent choices of particles $x_1,x_2$ nearest to the central
particle $x$ (in this symmetric case, $\rho_1=\rho_2$): along the
cube edge, along a face diagonal, and along a main diagonal of the
cube. Therefore, in the vicinity of this symmetric configuration,
the distance $\rho'_2=\|x_2-x_1\|$ cannot be a continuous function
of $k$-germ ($k\geq 4$).

Statistical theory of shapes of finite sets in $R^3$ was launched
in 1970s (see a survey \cite{Kend}).  Statistical analysis of
configuration germs is an interdisciplinary area between
statistics of shapes and statistical physics.

\subsection{Dynamics in systems with strong order--disorder
separation} \label{dynaccel}

In previous subsections we discussed relations between measure and
distance in high--dimensional systems with permutational symmetry.
But the main property of the measure under consideration is its
{\it invariance} with respect to mechanical motion. In this
subsection we consider dynamics in phase spaces with
concentration. Without such a return to dynamics the consideration
of order--disorder relations in statistical mechanics is
incomplete, and we can loose some important effects.

The strong order--disorder separation causes very special
peculiarities of dynamical systems with conservation of measure.
Let $P_n$ be the $n$-particle phase space for a system with strong
order--disorder separation. Assume that for given $\delta >0$ the
radius $\rho(\delta,n)$ of a $(1-{\delta})$-concentration ball
$B^{n \: {\rm conc}}_{\rho(\delta,n)}$ with the measure $1-\delta$
goes to 0 at $n \rightarrow \infty$ and the diameter of $P_n$ is
bounded: $\alpha \geq \mbox{Diam} P_n \geq \beta
>0$.\footnote{We consider compact spaces to avoid trivial technical
complications that are needed for locally compact spaces.} Phase
flow transformations form a one--dimensional semigroup of
injective maps $T_t: P_n \rightarrow P_n$, $T_t$ ($t>0$) is a
shift over time $t$. For any $t>0$ the map $T_t$ keeps the most
part of the $1-{\delta}$-concentration ball $B^{n \: {\rm
conc}}_{\rho(\delta,n)}$ in it:
\begin{equation}\label{intersection}
\mathbf{P}(T_t (B^{n \: {\rm conc}}_{\rho(\delta,n)}) \cap B^{n \:
{\rm conc}}_{\rho(\delta,n)}) \geq 1-2\delta,
\end{equation}
because the measure of complement of $B^{n \: {\rm
conc}}_{\rho(\delta,n)}$ in $P_n$ is less than $\delta$.

For time averages  of bounded differentiable vector--functions on
$[0, \infty[$ (with bounded derivatives) an elementary identity
holds:
\begin{equation}
\langle \dot{f}^2(t)\rangle = - \langle (f(t),\ddot{f}(t))\rangle,
\end{equation}
 if all averages exist, hence,
\begin{equation}
\langle \dot{f}^2(t)\rangle \leq \langle f^2(t) \rangle^{1/2}
\langle \ddot{f}^2(t))\rangle^{1/2},
\end{equation}
and
\begin{equation}
\frac{\langle \ddot{f}^2(t))\rangle}{\langle \dot{f}^2(t)\rangle}
\geq \frac{\langle \dot{f}^2(t)\rangle}{\langle f^2(t)\rangle}.
\end{equation}
Let us choose the origin in the center of  $B^{n \: {\rm
conc}}_{\rho(\delta,n)}$. In this case, under standard
assumptions,
\begin{equation}
\langle a^2\rangle \geq \frac{\langle
v^2\rangle^2}{\rho^2(\delta,n)},
\end{equation}
where $\langle a^2\rangle$ is the average square of acceleration,
$\langle v^2\rangle$ is the average square of velocity.

It means that in systems with concentration for given average
square of velocity $\langle v^2\rangle$ the average square of
acceleration tends to $\infty$ with the number of particles, even
if the velocity ($n$-particle) remains normalized. Just to imagine
the orders let us assume: ${\langle v^2\rangle} \sim
\frac{1}{2}nkT$, ${\rho^2(\delta,n)} \sim n^{-1}$. In this case,
$\langle a^2\rangle \geq const \times n^3$.

Dynamics of particles with elastic collisions on an interval   is
equivalent to billiards in a multidimensional simplex (see
elsewhere, for example \cite{sinai}). Even if the particles are
transparent (if there are no physical collisions at all), the
symmetric representation of the system (by a point in the symplex)
evolves with velocity jumps. These jumps take place every time
when the particles change their order on the line.

For a functional representation of moving particles
(\ref{funcrepr}) (for any dimension of the one-particle space) the
time derivative of $f_{x_1,x_2, \ldots , x_n}(x)$ has a
discontinuity when the nearest to $x$ particle changes its number.

In all these cases the motion in symmetric coordinates is only
piecewise differentiable, and average square of acceleration does
not exist at all (is infinite).

The described {\it acceleration--dominated dynamics} makes no
differences between real physical interaction and jumps of
velocities caused by geometry of permutation symmetry, for example.
For motion of particles on a line, the particles can be transparent
and do not interact at all. In this case one particle will come
through the other, but any change of their order on a line causes in
symmetric representation jump of velocities. On the other hand,
particles can interact, collide, and do not change their order on a
line. The result will be the same. For instantaneous elastic
collisions the difference does not exist,  but for softer potentials
the picture of acceleration dominance holds also. The system without
interaction is a billiard in the standard $n$-dimensional symplex.
Interaction changes (smoothes) collisions and bends trajectories
between them.

\section{Sticky faces and natural selection}

In this section we discuss general dynamical systems, not
necessarily  Hamiltonian ones, or systems with conservation of
volume. Let a multidimensional symplex be positively invariant
with respect to dynamics: if a motion start in this symplex at
some time $t_0$ then it belongs to the symplex at any moment
$t>t_0$. For such a dynamical system we can guess that the motion
spends most of time in a small vicinity of the symplex centre. It
is a very natural expectation because of the concentration of the
symplex volume near its centre, and some theorems in the form
``for a typical dynamical system with positively invariant
multidimensional symplex a typical motion spends most of the time
in a small vicinity of the symplex centre" could be proved for the
appropriate definition of typicallness. But there exists an
important opposite type of dynamic behaviour. Let us assume that
the faces of the symplex are also positively invariant. In this
case, the typical picture of dynamic behaviour changes
drastically: motions tend to a small vicinity of the
small--dimensional skeleton of the symplex.

Let us first explain the sense of such ``sticky faces." The
standard symplex $\Delta_n$ has natural interpretation as a space
of $n$-dimensional probability distributions $p_1,\ldots, p_n$
defined on $n$ states. A dynamic system with positively invariant
$\Delta_n$ is a kinetic equation. The faces of $\Delta_n$ are
positively invariant, if the $i$th state could not be produced
from the $j$th one for $i \neq j $, and only the birth--death rate
of $i$th state depends on the whole distribution $p_1,\ldots,
p_n$. It is the general form of {\it inheritance} property, and
such dynamical systems are standard objects for study in
mathematical biology after Volterra \cite{Volterra}, Lottka, and
Gause \cite{Gause}; review of some modern works could be found in
\cite{Odo}.

The concentration of motions for $t\rightarrow \infty$ in a small
vicinity of the small--dimensional skeleton of the symplex is
exactly the phenomenon of {\it natural selection} \cite{G1,GroGo}.
Many physical application of this phenomenon are known
\cite{Zakharov1,Zakharov2,Lvov}.

It is easy to demonstrate this phenomenon on the example of $n$
particles moving on an interval $[0,1]$. The effect of sticky faces
implies here that if the position of the $i$th particle is 0 or 1,
then it does not moves. Following natural hypothesis of smoothness
we can extract a multiplier $x_i$ near 0 and $(1-x_i)$ near 1 from
the velocity of the particle at the position $x_i$. It means that in
new coordinates $y_i=\ln x_i-\ln (1-x_i)$ we can expect more or less
uniform distribution of particles. But it is the equidistribution on
the whole line. It is impossible in the classical sense, but if we
take it seriously, we come to a finite--additive distribution (or to
an approximation with equidistributions on a sequence of extended
intervals). In any case, the expected number of particles at a given
distance from the interval ends (or, in $y_i$ coordinates, at a
given bounded interval) should be small in comparison with the total
number of particles: almost all particles are concentrated near
interval ends.

The whole effect of sticky faces in a simplex means that if some
coordinates $s_i$ are zero then their time derivatives are also
zero. For particles moving on $[0,1]$ it implies that they stick to
each other, and for such a system we can observe particle
agglutination, in addition to particle concentration near the
interval ends.

In order to achieve exact estimation and theorems it is useful to
start with an infinite number of particles. A variant of such a
theory for continuous families of particles is developed in
\cite{G1}, see English version in \cite{EvPrep}. The main result
remains the same: for the dynamics on a symplex with sticky
boundaries, almost all motions tend to a small vicinity of the
small--dimensional skeleton of the symplex. Estimates of the
skeleton dimension and asymptotic expansion for motions near this
skeleton are also obtained.

\section{Discussion}

For a large number of particles the available phase space (or
configuration space) could be divided into two parts. One part has
microscopically small diameter (part $D$, disorder), another part
(part $O$, order) has microscopically small measure (volume). This
is the strong {\it order--disorder separation}.

Permutation invariance is crucial for the strong order--disorder
separation. For example, the volume of a high--dimensional cube is
concentrated near its boundary. After symmetrization the cube
transforms into a simplex, and the volume of the simplex is
concentrated near its center. Order is in the long, but thin wings
of the simplex, while disorder is in the small, but thick vicinity
of its center. This effect allows many generalizations for spaces of
permutation orbits.

All individual configuration of $n$ distinguishable particles in
$R^m$ are equivalent: the measures of their
$\varepsilon$-vicinities coincide, and are equal just to a volume
of $nm$-dimensional ball of radius $\varepsilon$. The permutation
symmetry enforced us to replace any single configuration of $n$
particles $x$ by the set ${\bf S}_n x$  of $n!$ configurations
that are generated from $x$ by particles permutations.  These
finite ensembles are already not equivalent. For a given bounded
domain $P\subset R^m$ (a box) and large $n$, there exists such a
configuration $x_0$ of $n$ particles (``almost equidistribution")
in $P$ that $\varepsilon$-fattening of ${\bf S}x_0$, $\{{\bf S}_n
x_0\}_{\varepsilon}$, has ``almost all" volume of the
configuration space $P^n$: $${\rm Vol}(\{{\bf S}_n
x_0\}_{\varepsilon})/{\rm Vol}(P^m)>1-\delta$$ for $\varepsilon
\sim n^{-1/m}$, and a given small number $\delta$. If such a
configuration exists, then, obviously, all points $x$ from $\{{\bf
S}_n  x_0\}_{\varepsilon}$ have the same property (with twice
increased $\varepsilon$): $${\rm Vol}(\{{\bf S}_n
x\}_{2\varepsilon})/{\rm Vol}(P^m)>1-\delta .$$ This finite
ensemble ${\bf S}_n x$ is not an $\varepsilon$-net in $P^n$,
moreover, the rest,  $P^n \setminus \{{\bf S}_n
x\}_{\varepsilon}$, has macroscopic (non-small) diameter and
macroscopic Hausdorff distance from ${\bf S}_n x$. This is the
essence of the strong order--disorder separation: disorder has
microscopic diameter (but macroscopic, almost all measure), order
has microscopic measure (but macroscopic diameter).

The disordered states from  the set $D$ are macroscopically
indistinguishable, because the distance between them is
microscopically small. We can use a notion of ``observability"
from the control theory, and say that the difference between these
states is macroscopically inobservable.

In our definition of order and disorder we use the state--based
approach: a state may be ordered or disordered. (Of course, in the
definition of order we use $\varepsilon$-fattening (\ref{fatt}),
hence, an ensemble is present too, but the notion of order relates
to states.) The state--based point of view in foundation of
statistical physics becomes more popular very recently
\cite{GorOrdPrep,GoLeTuZa06}.

The time arrow that leads from order to disorder has the following
interpretation: if a motion starts from an ordered state, then,
after some time, the state becomes disordered, and we can be
almost sure that it will remain disordered during time $T$ with
microscopically small inverse $T^{-1}$ (probability of fluctuation
from  disorder to order could be estimated on the basis of
Eq.~(\ref{intersection})). Neither chaotic dynamics, no dynamical
stirring have any relation to this behaviour. Even ergodicity is
not especially important: if ergodic components are
multi-particle, then the same order-disorder separation is
expected on them. But without strong order--disorder separation it
is impossible even to formulate such a statement: if a motion
starts from an ordered state, then, after some time, the state
becomes disordered, and ...

Dynamics of systems with strong order--disorder separation has a
very special property: in symmetric representation the average
square of acceleration is very high. It can be interpreted as
evolution through a series of collisions even for non-interacting
particles. It is a hint to a possible solution of an essential
open problem, the {\it problem of indivisible events}. For a
macroscopically small time, a small microscopic subsystems can go
through ``its whole life", from the beginning to the limit state
(or, more accurate, to the limit behaviour which may be not only a
state, but a type of motion, etc.). The evolution of the
microscopic subsystems in a macroscopically small time $\Delta t$
should be described as an {\it ``ensemble of indivisible events"}.
An excellent hint is given by the Boltzmann equation with its
indivisible collisions, another good hint gives the chemical
kinetics with indivisible events of elementary reactions. Now we
understand that the solution could be found in the high
acceleration for systems with strong order--disorder separation
(Subsec.~\ref{dynaccel}), but don't know yet even a form of a
proper answer.

The effect of strong order--disorder separation and time arrow
direction from order to disorder turn to inverse, if we assume
invariance of the boundary (sticky boundaries). If we consider
these dynamical systems as kinetic equations, the effect of sticky
boundaries can be presented as inheritance: if some species (or
genes -- for our choice) are not present in the system now, they
will not appear in the future. In this case, the evolution from
disorder to order has a special name: Natural selection. Many
applications of this effect are known in physics: from mode
selection in lasers to wave turbulence. It is as general as
order--disorder separation, and appears together with any sort of
inheritance.

The role of the permutation symmetry in statistical physics was
discussed many times, from different points of view: as a basic
axiom \cite{Cost,SantSant}, as a practical question related to
entropy definition and measurement \cite{Blum,Nagl}; even an
ontological status of this assumption was discussed quite
thoroughly \cite{Gordon}. In this paper, in addition to this
discussion, we demonstrate importance of permutation invariance
for order--disorder separation and for direction of time arrow
from order to disorder.

The idea of measure concentration already affects even applied
computer science \cite{Pestov}. The history of physical
applications starts more than 100 years ago, and now the measure
concentration is one of the central ideas of statistical physics,
we should only recognize  this properly.

\begin{center}
* * *
\end{center}

And what about the children-- and parents--rooms? Of course, the
children--room has no permutation symmetry: any toy has it own
sense, and a permutation destroys the sense of the configuration.
But in parents--room there is perfect permutation symmetry: the
toys there are just some things that should be returned in the
toy--box. Hence, the children room is in order, but the parents
room is in full disorder. Moreover, the children--room has no
order--disorder separation, because each configuration has its own
sense: the disorder is impossible in the children--room!

{\bf Acknowledgements}. I am very grateful to M.~Gromov and
H.C.~\"Ottinger for inspiring discussions, and to R.~Davidchack
with A.~Zinovyev whose first computer experiments encourage me to
publish here recommendations about local statistical analysis of
particle configuration.

\end{document}